\documentclass{Interspeech}

\interspeechcameraready

\title{Personalized Voice Synthesis through Human-in-the-Loop Coordinate Descent}

\author[affiliation={1}]{Yusheng}{Tian}
\author[affiliation={1}]{Junbin}{Liu}
\author[affiliation={1}]{Tan}{Lee}

\affiliation{Department of Electronic Engineering}{The Chinese University of Hong Kong}{Hong Kong SAR}
\email{ystian0617@link.cuhk.edu.hk, liujunbin@link.cuhk.edu.hk, tanlee@ee.cuhk.edu.hk}

\usepackage{comment}
\usepackage{cite}
\usepackage{algorithm}
\usepackage{algpseudocode}
\usepackage{array}
\usepackage{setspace}
\usepackage{anyfontsize}

\begin{document}

\maketitle

\begin{abstract}
    
    
    This paper describes a human-in-the-loop approach to personalized voice synthesis in the absence of reference speech data from the target speaker. It is intended to help vocally disabled individuals restore their lost voices without requiring any prior recordings. The proposed approach leverages a learned speaker embedding space. Starting from an initial voice, users iteratively refine the speaker embedding parameters through a coordinate descent-like process, guided by auditory perception. By analyzing the latent space, it is noted that that the embedding parameters correspond to perceptual voice attributes, including pitch, vocal tension, brightness, and nasality, making the search process intuitive. Computer simulations and real-world user studies demonstrate that the proposed approach is effective in approximating target voices across a diverse range of test cases.
\end{abstract}

\section{Introduction}
Personalized voice synthesis offers a powerful tool for individuals with vocal disabilities to restore their lost voices\cite{veaux2012using, erro2015personalized, cave2021voicebank}. By adapting or prompting an existing speech synthesis model with speech recordings of a target person, a synthetic voice that closely resembles the original can be created \cite{voiceclone,svtotts,adaspeech4,naturalspeech3}. The personalized synthetic voice can be used to generate speech of new content, allowing the person to express thoughts and feelings in a way that feels uniquely his/her own. 
However, this approach relies on the availability of reference recordings, excluding those people who lost their voices before the recordings could be made, e.g., head and neck cancer survivors.

The present study was initiated with a simple and unpretentious purpose: to help voiceless individuals regain the ability of producing speech in their own voices in the absence of reference recordings. When physical recordings are not available, the only trace of the voices resides in the auditory memory of relevant human listeners. This presents a unique challenge: synthesizing a target voice that the user can clearly identify but for which no reference recordings exist. 


The task we are considering is essentially a black-box optimization problem, where the goal is to maximize the perceived similarity between a computer-generated voice and what the user has in mind. The objective function, depending on the user’s subjective perception, is unknown and cannot be directly measured. This mirrors the task of sound design\cite{sounddesign}, where musicians craft desired audio effects by adjusting parameters in a synthesizer.

In this paper, we introduce a voice synthesizer that enables human listeners to shape voice characteristics in a way similar to sound design. The system is built on a general speech resynthesis framework, in which the learned speaker embeddings capture information pertinent to voice characterization, e.g., pitch, timbre. Principal Component Analysis (PCA) is applied to transform the embeddings into a concise set of parameters that can be managed by users. Starting with an input utterance as the initial voice, users can iteratively refine salient voice characteristics through focused listening. 

To guide users toward their target voices, an interactive search algorithm is designed and implemented. The search process resembles the classic coordinate descent, with each iteration exploring a single principal component direction, guided by user feedback during a listening and comparison task. Post-hoc analysis shows that these directions correspond to distinct voice qualities, such as pitch, vocal tension, brightness, and nasality. This connection between adjustable parameters and audible characteristics makes the search process intuitive, effectively reducing the cognitive load on the user. 

In summary, our main contribution is a human-in-the-loop approach to synthesizing target voices based on user feedback. This approach is particularly valuable for speech-impaired individuals who wish to recreate their lost voices but lack prior recordings. By completing a series of listening and comparison tasks, users can progressively refine the system’s output to match their target voices in an intuitive way. Both computer simulations and real-world user studies demonstrate that the proposed search algorithm can approximate target voices across a diverse range of test cases.

\section{Related work}
Recent research has explored using textual descriptions as an alternative to exemplar recordings for customized voice synthesis \cite{prompttts,leng2024prompttts2,prompttts++,promptspeaker,vyas2023audiobox,hai2024dreamvoice, lyth2024natural}. For example, prompts like ``a low-pitched female voice with a dark tone" can guide the synthesis process. However, text-based methods alone struggle to capture highly specific target voices. This limitation stems from the subjective nature of voice perception and the inherent inadequacy of describing voice qualities in words. For instance, one person's idea of a ``warm" voice can differ substantially from another's, and a single textual description may apply to many distinct voices.

Another line of research has explored using human feedback to tailor synthesized voices for specific contexts \cite{voiceme,robotvoice,udagawa22_interspeech, hilvoice}, such as matching a given face or appealing to the elderly audience group. This human-in-the-loop strategy directly inspired our work. In these studies, listeners refine synthesized voices by adjusting speaker embedding parameters of a pre-trained TTS model, thus integrating human perception into the voice generation process. 
However, TTS-derived speaker embeddings often entangle multiple attributes such as timbre, accent and speaking style. Adjusting one parameter can unintentionally affect others, making the refinement process inefficient and cognitively demanding. 
While these methods can generate voices that are contextually plausible and align with general human preferences, they lack the precision needed to recreate a target voice.
We address this issue by constructing the speaker embedding space within a speech resynthesis framework, which is shown to produce more interpretable parameters that correspond to distinct voice qualities. 

\section{Method}
\subsection{System overview}
\begin{figure}[t]
  \centering
  \includegraphics[width=0.8\linewidth]{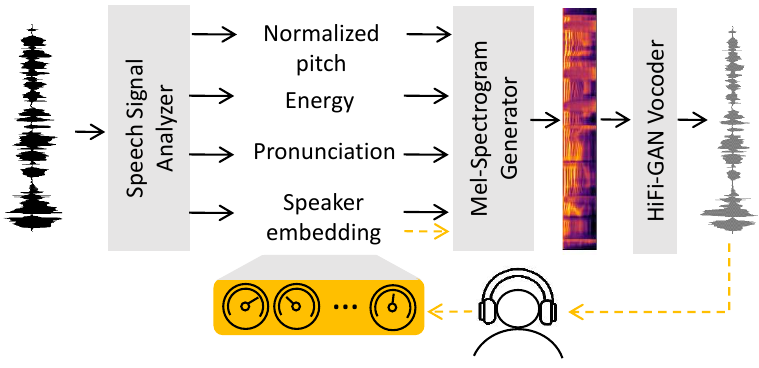}
  \caption{Overview of the proposed system. Per-utterance pitch normalization is applied to encode both the timbre and overall pitch information into the speaker embedding.}
  \label{fig:overview}
  \vskip -0.2in
\end{figure}

The proposed system is built on a speech resynthesis framework similar to NANSY \cite{nansy}. As illustrated in Figure~\ref{fig:overview}, it decomposes any input speech signal into four key features---pitch, energy, pronunciation, and speaker representation---then recombines them to either reconstruct the original utterance or produce a modified one. We treat PCA-reduced speaker embeddings extracted with this framework as the parametric representation of voice. Since pitch is essential in characterizing a person's voice, per-utterance pitch normalization is applied so that pitch information is also encoded in the speaker embedding. 

With an input utterance as the initial voice, users can adjust parameters of the speaker embedding to customize the overall pitch and voice timbre without altering other speech factors, effectively performing voice conversion. While our primary focus is on pitch and timbre, the modular architecture makes it convenient to introduce additional controls for other speech dimensions, such as accent and speaking style.

\subsection{Human-in-the-loop search algorithm}
By applying PCA to speaker embeddings extracted from a large multi-speaker speech corpus, we obtain a set of principal directions. Any speaker embedding $\boldsymbol{z}$ can be approximated by projecting it onto the subspace spanned by the top $N$ principal directions, expressed as $\boldsymbol{z} \approx \boldsymbol{\mathrm{W}}\boldsymbol{\alpha}+\boldsymbol{\mathrm{b}}$, where $\boldsymbol{\alpha}=[\alpha_1,\alpha_2,\dots, \alpha_N]^T$ are the linear combination coefficients, $\boldsymbol{\mathrm{W}}=[\boldsymbol{\mathrm{w}}_1, \boldsymbol{\mathrm{w}}_2,\dots, \boldsymbol{\mathrm{w}}_N]$ is the basis matrix, and $\boldsymbol{\mathrm{b}}$ is an offset vector. By fixing $\boldsymbol{\mathrm{b}}$ to a predefined value (e.g. the average speaker embedding), we reduce the problem to optimizing the $N$ coefficients $\boldsymbol{\alpha}$ to maximize the user's perceptual preference for the generated voice:
\setlength{\abovedisplayskip}{6pt} 
\setlength{\belowdisplayskip}{6pt} 
\begin{equation}
\label{eq:optimization}
\max_{\boldsymbol{\alpha}\in\mathbb{R}^N} {f\left( g\left({\boldsymbol{\mathrm{W}}\boldsymbol{\alpha}+\boldsymbol{\mathrm{b}}}\right)\right)},
\end{equation}
where $g(\cdot)$ represents the voice synthesizer, and $f(\cdot)$ denotes the user's perceptual preference function, respectively. Features that remain unchanged, such as energy and pronunciation sequences, are omitted here for conciseness.

The objective function in (\ref{eq:optimization}) includes a perceptual judgment that cannot be directly measured. We therefore employ a human-in-the-loop approach. It follows the classic coordinate descent \cite{wright2015coordinate}, optimizing one principal component coefficient at a time while keeping the others fixed, and cycling through all parameters until convergence. Unlike the classic coordinate descent, each one-dimensional step is guided by user feedback to incorporate human perception into the search process.

As shown in Algorithm~\ref{alg:cap}, the search process starts with an initial speaker embedding $\boldsymbol{z}^{(0)}$. To explore variations, we systematically perturb $\boldsymbol{z}^{(0)}$ along the first principal direction, creating a set of candidate speaker embeddings. Each candidate is then synthesized into an audio sample. The user listens to these audio samples and selects the most similar voice. The chosen embedding then becomes the starting point for the next iteration, which explores variations along the next principal direction. This cycle continues, traversing all principal directions repeatedly until the user finds a satisfactory voice or a predefined maximum number of iterations is reached.

\begin{algorithm}
\caption{Human-in-the-loop search algorithm}\label{alg:cap}
{\footnotesize
\setstretch{0.9}
\begin{algorithmic}
\Require $N$ principal directions $\boldsymbol{\mathrm{w}}_1, \boldsymbol{\mathrm{w}}_2,\dots, \boldsymbol{\mathrm{w}}_N$
\Require $N$ step sizes $d_1, d_2, \dots, d_N$
\Require Initialization $\boldsymbol{z}^{(0)}$
\State $i \gets 0$
\While{stopping criteria is not met}
\State $n \gets i\mod{N}+1$ \Comment{principal component index}
\For{$k=\pm1,\pm2,0$}
\State $\boldsymbol{z}_{k}^{(i)} \gets \boldsymbol{z}^{(i)}+k\cdot2^{-\lfloor\frac{i}{N}\rceil}d_n\boldsymbol{\mathrm{w}}_n$ \Comment candidate voices
\EndFor
\State $\boldsymbol{z}^{(i)}_{\textrm{sel}} \gets \arg\max { f\left( g\left({\boldsymbol{z}_k^{(i)}}\right)\right)}$ 
\Comment{{user response}}
\State $\boldsymbol{z}^{(i+1)}\gets \boldsymbol{z}_{{\textrm{sel}}}^{(i)}$, $i\gets i+1$
\EndWhile
\end{algorithmic}
}
\end{algorithm}
\vskip -0.1in

\begin{figure}[!t]
\includegraphics[width=0.97\linewidth]{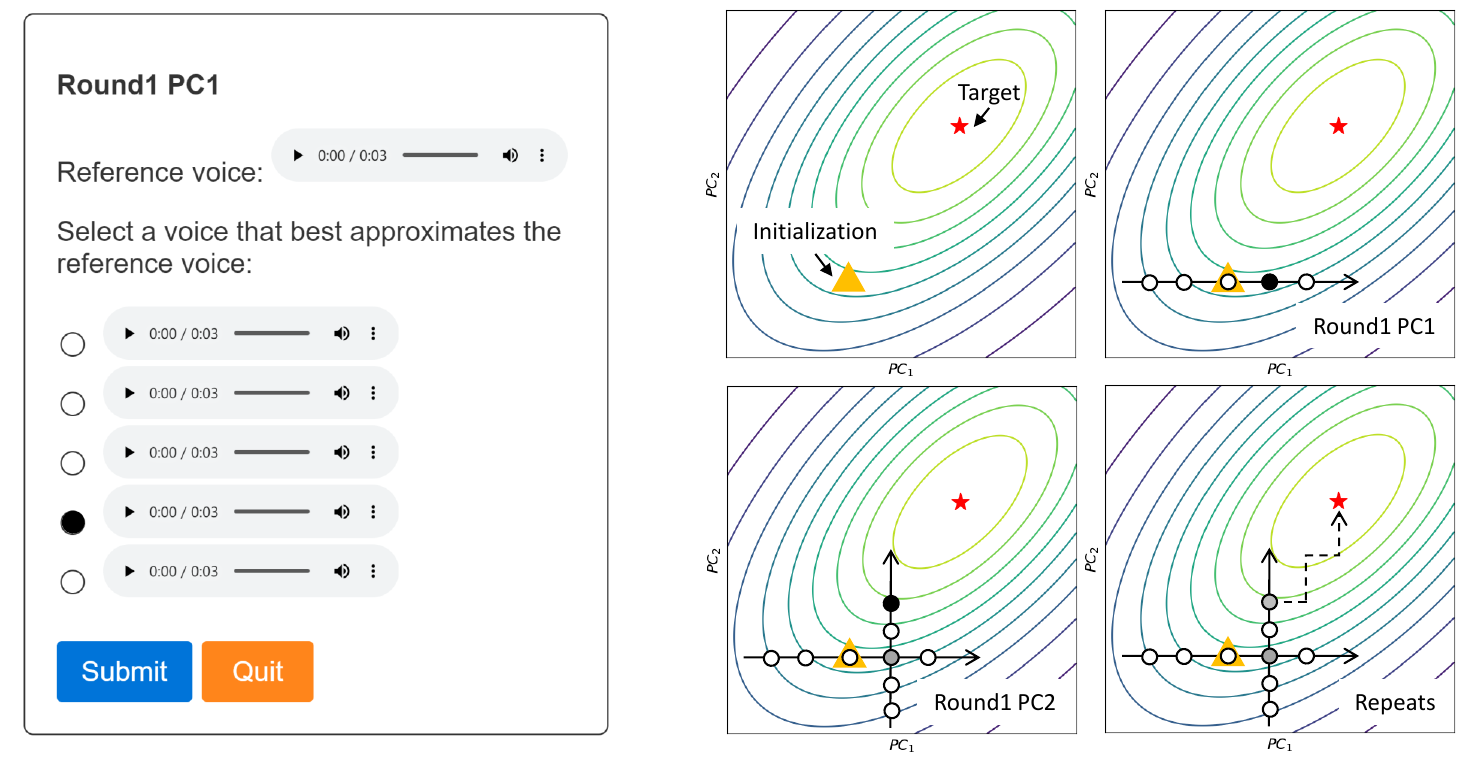}
\centering
\caption{Illustration of the search process. Left: the user interface for a single query. Right: an example search sequence within a 2-dimensional parameter space. }
\label{fig:user_interface}
\vskip -0.2in
\end{figure}

Figure~\ref{fig:user_interface} illustrates how user feedback drives the search process. For evaluation purposes, the interface provides a reference voice, and participants are asked to select the candidate voice most similar to it. This setup allows us to assess how closely the search result matches the target voice. In practical use, however, no reference is provided. Users are expected to rely on their mental image of the target voice to make their choices. 

\section{Experimental setup}
\subsection{Model and dataset}
Given an input utterance, the speech signal analyzer in Figure~\ref{fig:overview} extracts pitch values using the DIO algorithm \cite{dio}. These values are normalized to have zero-mean and standard deviation, then encoded into 128-dimensional vectors.
Energy features are computed by summing the log mel-spectrogram along the frequency axis, followed by the same encoding process.
Speaker embeddings are derived using the ECAPA-TDNN network \cite{ecapa-tdnn}, which produces 192-dimensional speaker embedding vectors.
Pronunciation features are extracted with ContentVec\cite{contentvec}, using the publicly released pretrained checkpoint.

The encoded pitch, energy and pronunciation features are summed and fed to a mel-spectrogram generator, which consists of six feed-forward transformer blocks.
The first four blocks take the ${\ell}_2$-normalized speaker embedding to control the layer normalization for both self-attention and feed-forward layers. The generator's output is then passed to a pretrained HiFi-GAN vocoder\footnote{https://github.com/jik876/hifi-gan} to produce the final speech waveform.

The network is trained on the LibriTTS-R training subsets \cite{libritts-r}. 
Only utterances longer than 2.0 seconds are used, corresponding to 527-hour speech data from 2,293 speakers. Model parameters are updated for 200,000 steps using the Adam optimizer at a learning rate of 1e-4 and a batch size of 32.

\subsection{Principal Component Analysis of speaker embeddings}
We use the trained ECAPA-TDNN network to extract a speaker embedding from each training utterance, then average these embeddings per speaker, yielding 2,293 distinct 192-dimensional vectors. We apply PCA to these vectors separately for female and male voices. The top 16 principal components capture over 75\% of the total variance.

To determine how many principle components are sufficient to capture key voice characteristics, we projected the extracted speaker embeddings onto a reduced set of principal components and evaluated the fidelity of voices reconstructed from these reduced representations. Specifically, for each of the training speakers, we randomly selected one utterance and generated two reconstructed voices: one using the full-dimensional embedding ($\boldsymbol{z}$) and another using a reduced-dimensional embedding ($\hat{\boldsymbol{z}}$). The reduced embedding was obtained using the formula: $\hat{\boldsymbol{z}}\gets\boldsymbol{\mathrm{W}}_K\boldsymbol{\mathrm{W}}^T_K(\boldsymbol{z}-\boldsymbol{\mu})+\boldsymbol{\mu}$, where $\boldsymbol{\mathrm{W}}_K$ represents the matrix containing the first $K$ principal components, and $\boldsymbol{\mu}$ is the average speaker embedding. We then measured the similarity between the two reconstructed voices (full-dimension vs. reduced-dimension) using Resemblyzer\footnote{{https://github.com/resemble-ai/Resemblyzer}}, a widely adopted open source speaker encoder. Our results indicate that using $16$ components is sufficient to achieve a Resemblyzer similarity score above $0.81$ for most training voices. This threshold corresponds to the $75^{\textrm{th}}$ percentile of intra-speaker similarity on the training data, representing a reasonably high similarity. We also noticed that differences beyond the $16^{\text{th}}$ component are imperceptible to human listeners. Therefore, we set $N=16$ in Algorithm~\ref{alg:cap} for subsequent experiments.

To highlight the strengths of using a speech resynthesis framework to construct the speaker embedding space, we compared our learned embeddings with those obtained from a VITS TTS model \cite{vits}. We trained VITS on the same LibriTTS-R dataset using the publicly available script\footnote{{https://github.com/jaywalnut310/vits}} and then performed PCA on the resulting speaker embeddings. We observed a key difference: the embeddings from our framework exhibited a more structured organization. Specifically, the first few principal components of our embeddings primarily corresponded to pitch variations. In contrast, manipulating VITS-derived embeddings along most principal component directions affected not only pitch but also other voice attributes. This organized structure in our framework facilitates more intuitive user exploration of the embedding space. Section \ref{sec: interpretation} provides detailed analysis of how each principal component correlates with perceptual attributes. We also encourage readers to explore these differences interactively on our demo page\footnote{https://myspeechprojects.github.io/voice-design-demo/}.

\section{Computer simulation}
To evaluate the algorithm's performance across diverse test cases, we adopt a simulation-based approach with a surrogate objective function. This computer simulation circumvents the prohibitive costs of human-in-the-loop testing and helps identify the most representative samples for the later user study.

We propose a heuristic surrogate function with two terms to approximate human perception of voice similarity. The first term is the voice similarity score measured by Resemblyzer\footnote{{https://github.com/resemble-ai/Resemblyzer}}, a widely adopted open-source speaker encoder. The second term is the Mean Squared Error (MSE) of the mel-spectrogram. 
Given a reference speech signal $\boldsymbol{y}^{\textrm{ref}}$ and a speaker embedding $\boldsymbol{z}$ under evaluation, the surrogate objective value is computed as
\begin{equation}
\label{eq:surrogate}
S(\boldsymbol{z})= \textrm{SimScore}\left(\boldsymbol{y}_\textrm{wav}^{{\boldsymbol{z}}},\, \boldsymbol{y}_\textrm{wav}^{\textrm{ref}}\right) -\textrm{MSE}\left(\boldsymbol{y}_\textrm{mel}^{{\boldsymbol{z}}},\, \boldsymbol{y}_\textrm{mel}^{\textrm{ref}}\right),
\end{equation}
where $\boldsymbol{y}_{\textrm{mel}}^{{\boldsymbol{z}}}$ is the reconstructed mel-spectrogram, obtained by replacing the original speaker embedding with $\boldsymbol{z}$ while preserving all other speech features, and $\boldsymbol{y}_{\textrm{wav}}^{\boldsymbol{z}}$ is the corresponding speech waveform synthesized using the HiFi-GAN vocoder. 

To account for variability in human preferences, we add Gaussian noise to the surrogate objective. When comparing two candidate embeddings $\boldsymbol{z}_1$ and $\boldsymbol{z}_2$, we assume the user prefers the first if
\begin{equation}
    S(\boldsymbol{z}_1) + \epsilon_1 > S(\boldsymbol{z}_2) + \epsilon_2, 
\end{equation}
where $\epsilon_1$ and $\epsilon_2$ are independent Gaussian variables with a standard deviation of 0.01.

The simulation executes the search process described in Algorithm 1, using the surrogate objective function to mimic user responses. The search process is set to terminate after 32 queries to reflect the limitation of real user patience. We evaluate the algorithm on test samples drawn from the ``test-clean" and ``test-other" subsets of LibriTTS-R and the VCTK \cite{vctk} dataset. We randomly select one utterance per speaker, yielding 72 unique target voices from LibriTTS-R and 110 from VCTK. In each simulation, we explore $N=16$ principal direction. The step size along the $i^{\text{th}}$ direction is set to $d_i=\sigma_i$, the standard deviation of the projected speaker embeddings. The simulation is repeated 20 times for each test sample, with each run starting from a randomly selected training-set voice.

%
\begin{table}[t]
\caption{Success rates (\%) for the LibriTTS-R and VCTK test samples, computed across 20 different initializations.}
\centering
{\footnotesize
\setstretch{0.87}
\label{tab:successrate}
\begin{tabular}{lccccr}
\toprule
 & \textbf{Mean $\pm$ Std}& \textbf{Max} & \textbf{Min} \\
\midrule
LibriTTS-R    & 97.7$\pm$ 5.2& 100.0 & 80.0 \\
VCTK & 87.5$\pm$ 18.1&100.0& 15.0\\
\bottomrule
\end{tabular}
}
\vskip -0.2in
\end{table}
We consider a search successful if, among the 32 queries, at least one selected candidate achieves a Resemblyzer score above 0.81. This threshold corresponds to the 75th percentile of intra-speaker similarity on the training data, representing a reasonably high similarity. Table~\ref{tab:successrate} summarizes the success rates for all test samples. The LibriTTS-R test set achieves a minimum success rate of 80\%, suggesting that if one initialization fails, trying another one is likely to succeed. However, results from the VCTK dataset show more variance in success rates. This indicates that, for out-of-domain voices, the algorithm's performance can be more sensitive to initial conditions.

\section{User study}
We conducted a user study to investigate the following questions: (1) Do human listeners succeed in finding good matches when the simulated search performs well? (2) Why does the simulated search struggle in certain scenarios, and would humans encounter similar difficulties in those situations? 

For this purpose, we selected two pairs of target voices, one pair from each of the two datasets used. Each pair consisted of an {``easy''} target and a {``hard''} target. The ``easy" target was defined as the one with the highest average Resemblyzer similarity score across the 20 simulation runs. Conversely, the ``hard" target was the one with the lowest average similarity score.

Five participants completed the user study via a web interface. All of them have passed a screening test demonstrating their ability to discern changes in voice produced by perturbing a speaker embedding along the 16 principal directions. Each participant was assigned all four target voices. For each target voice, they started from an initial voice known to produce a successful search in our simulation, and the maximum number of iterations was set to 32. Participants first completed a training trial to familiarize themselves with the procedure and user interface. In the formal trials, they had up to three attempts to search for each target voice, and the attempt they identified as the closest match was saved for evaluation.
\begin{figure}[!t]
\centering
\includegraphics[width=1.0\linewidth]{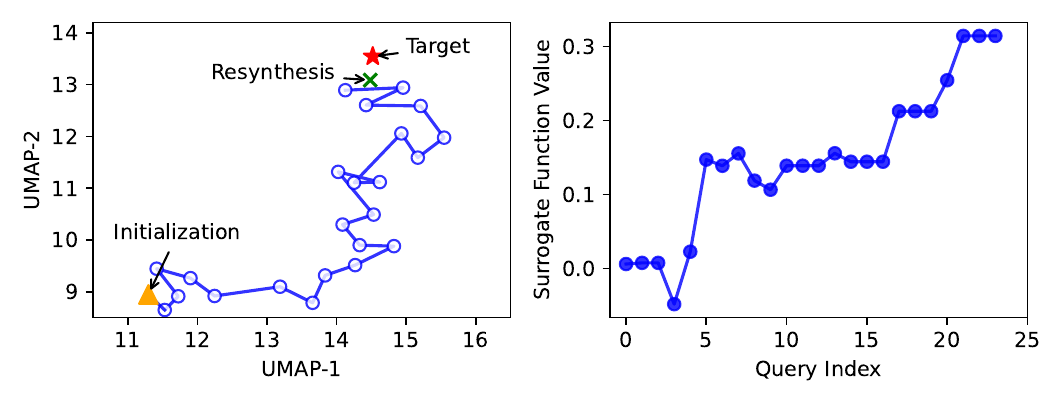}
\centering
\caption{A real user session for the LibriTTS-R ``easy" target. Left: UMAP-projected speaker embeddings, extracted using Resemblyzer. Right: Surrogate objective function value of user-selected voice candidates after each query.}
\label{fig:umap}
\end{figure}

Figure \ref{fig:umap} illustrates a real user session for the LibriTTS-R ``easy" target voice. Although the UMAP projection of the search trajectory and the surrogate objective function values show some fluctuations, the overall trend is towards increased similarity. This suggests a general agreement between the surrogate objective function and human perception of similarity.

We then conducted a MUSHRA-style listening test. For each of the target voices, listeners compared eight audio samples: those found by the five participants and the simulated search, the PCA-reconstructed voice using 16 principal components, and the initial voice. Six native English speakers rated the similarity of each sample to the reference original recording on a scale of 0-100, with scores above 60 considered ``good" and scores above 80 considered ``excellent".
 
 Table~\ref{tab:mushr} presents the mean and standard deviation of the listening test scores. As shown, participants in the user study generally achieved excellent or near-excellent scores on the ``easy" targets. Also, the gap between the ``easy" and ``hard" cases for the VCTK dataset is substantial, consistent with the earlier simulation results. Participants reported that reaching the ``hard" VCTK voice required more steps, suggesting that later principal components play a more significant role in these cases. This issue might stem from the limited presence of similar voices in the training data. Note that the PCA-reconstructed ``hard" VCTK voice scored just above 70, indicating that the constraint lies not in the search algorithm itself, but in the limited representation of out-of-domain voices. 
\newcolumntype{P}[1]{>{\centering\arraybackslash}p{#1}} 
\newcolumntype{L}[1]{>{\raggedright\arraybackslash}p{#1}} 

\begin{table}[!t]
\renewcommand{\arraystretch}{1.0}
\caption{Mean scores from the MUSHRA-style listening test.}
\label{tab:mushr}
\centering
{\footnotesize
\setstretch{0.87}
\setlength{\tabcolsep}{2pt} 
\begin{tabular}{@{}L{14mm} *{4}{P{14mm}}} 
\toprule
& \multicolumn{2}{c}{LibriTTS-R} & \multicolumn{2}{c}{VCTK} \\
 & \textbf{Easy} & \textbf{Hard} & \textbf{Easy} & \textbf{Hard} \\
 \midrule
User 1 & 77.8 $\pm$ 4.6 & 68.2 $\pm$ 4.9 & 83.9 $\pm$ 3.9 & 53.3 $\pm$ 2.5 \\
User 2 & 52.2 $\pm$ 7.8 & 80.6 $\pm$ 3.7 & 79.2 $\pm$ 3.3 & 62.8 $\pm$ 6.7 \\
User 3 & 70.3 $\pm$ 5.1 & 68.4 $\pm$ 7.2 & 79.0 $\pm$ 1.7 & 68.3 $\pm$ 5.7 \\
User 4 & 78.9 $\pm$ 3.8 & 76.4 $\pm$ 5.4 & 83.9 $\pm$ 2.6 & 64.0 $\pm$ 5.9 \\
User 5 & 79.4 $\pm$ 3.7 & 79.9 $\pm$ 5.5 & 84.5 $\pm$ 3.5 & 75.0 $\pm$ 3.8 \\
Simulation & 71.2 $\pm$ 4.1 & 71.5 $\pm$ 5.2 & 82.3 $\pm$ 2.9 & 68.5 $\pm$ 3.4 \\
PCA-recon & 85.6 $\pm$ 4.0 & 86.1 $\pm$ 3.9 & 82.9 $\pm$ 2.2 & 70.3 $\pm$ 2.6 \\
Initialization & 42.3 $\pm$ 4.0 & 46.6 $\pm$ 4.9 & 34.9 $\pm$ 4.1 & 38.3 $\pm$ 4.3 \\
\bottomrule
\end{tabular}
}
\vskip -0.2in
\end{table}

\section{Interpreting PCA directions} \label{sec: interpretation}
More than one participant in the user study noted that early adjustments mainly affect pitch, while later adjustments influence qualities like vocal strain and brightness. This observation suggests that certain directions in the speaker embedding space may correspond to specific voice attributes, potentially aligning with the PCA directions. To explore this idea, we adapt techniques from the computer vision domain to discover voice editing directions within the latent space, then compare these directions to the PCA directions.

Given a speech sample, let $\boldsymbol{z}$ be its speaker embedding and $\boldsymbol{x}$ be the other speech features extracted within our framework. A small perturbation of $\boldsymbol{z}$ in the direction of $\boldsymbol{v}$ changes the generated mel-spectrogram according to the following derivative:
\begin{equation}
\label{eq:derivative}
    \lim_{\epsilon\rightarrow 0} \frac{{G}_{\boldsymbol{x}}({\boldsymbol{z}+\epsilon\cdot\boldsymbol{v}})-{G}_{\boldsymbol{x}}({\boldsymbol{z}})}{\epsilon}=J_{{G}_{\boldsymbol{x}}}(\boldsymbol{z})\cdot\boldsymbol{v},
\end{equation}
where ${G}_{\boldsymbol{x}}(\boldsymbol{z})$ represents the mel-spectrogram generator function, and $J_{{G}_{\boldsymbol{x}}}(\boldsymbol{z})$ is its Jacobian evaluated at $\boldsymbol{z}$. 

One assumption in generative model latent space analysis, supported by several previous studies\cite{spectralregularizer,lowranksubspace,closedform,jacobianregularizer}, is that effective editing directions should induce the most significant changes in the generated output.
Guided by this principle, we examine directions that produce the largest changes in the generated mel-spectrogram, which correspond to the leading right singular vectors of the Jacobian $J_{{G}_{\boldsymbol{x}}}(\boldsymbol{z})$ in Equation
(\ref{eq:derivative}). Specifically, we selected one utterance from each speaker in the training data. For each utterance, we computed the associated Jacobian matrix, and extracted the top 16 right singular vectors to form a pool of candidate editing directions. To ensure that these directions generalize across speakers and speech content, we applied DBSCAN clustering\cite{dbscan}, using cosine similarity as the distance metric with a threshold of $0.1$. Clustering was performed separately for female and male voices. Interestingly, this yielded five distinct editing directions for each group.
\begin{figure}[!t]
\centering
\includegraphics[width=1.0\linewidth]{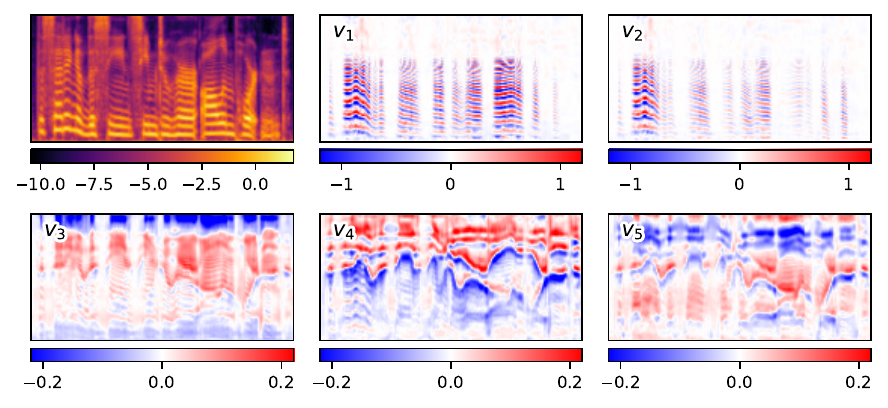}
\centering
\caption{
Visualization of how the generated mel-spectrogram changes when the speaker embedding is shifted along the five principal voice editing directions. The selected speech sample is 6345\_93302\_000037\_000003.wav from the LibriTTS-R dev-clean set, a female voice speaking ``The setting of the scene seemed to her all important". }
\label{fig:jacobian}
\vskip -0.2in
\end{figure}
%

To determine which voice attributes these directions control, we manually examined a few utterances and tentatively labeled the five directions as corresponding to pitch level, pitch variance, vocal strain, brightness, and nasality. Figure \ref{fig:jacobian} provides a concrete example, illustrating how the generated mel-spectrogram of a sample utterance changes when the speaker embedding is shifted along the six voice editing directions. 

For example, moving the speaker embedding along the first direction appears to erase the original horizontal stripes in the mel-spectrogram and redraw them at a slightly higher frequency. This suggests that this direction primarily controls the pitch level of the generated voice. The second direction also induces an upward shift in frequency bands but does so selectively, primarily affecting time windows where the fundamental frequency is already high. This targeted manipulation results in increased pitch variation rather than a uniform pitch shift. These two directions are also highly correlated with the top three principal directions discovered by PCA. Perturbing the speaker embedding along the third direction, concentrates added energy within the main body of the frequency bands, a change that corresponds to increased vocal tension or strain. The effects of the last two directions are less apparent from visual inspection of the gradient maps alone. However, upon listening to the manipulated utterances, we observe that the forth direction controls the level of nasality in the voice, while the fifth direction affects the overall timbre, shifting a bright voice towards a more muffled quality.

\begin{figure}[!t]
\includegraphics[width=0.87\linewidth]{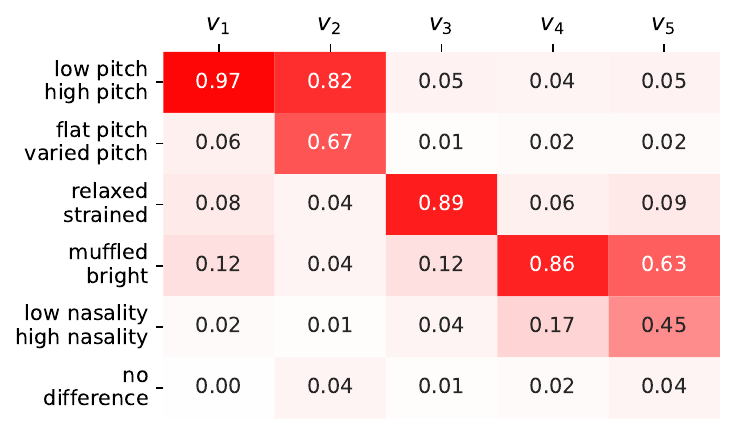}
\centering
\caption{Listener responses of their perceived changes in voice attributes when the speaker embedding is manipulated along a single editing direction.}
\label{fig:attribute_edit}
\vskip -0.2in
\end{figure}

Next, we developed a web-based listening test in which participants heard pairs of manipulated utterances: one with the speaker embedding shifted in a positive editing direction and the other in the opposite direction. For each pair, participants were asked to identify any noticeable changes among the five labeled attributes or to select ``no difference". In total, 360 unique audio pairs (5 directions $\times$ 72 utterances) were created from the LibriTTS-R test sets, and a random subset of 10 questions was presented in each survey. Since the listening test involves identifying subtle voice differences, we recruited participants via social media, targeting individuals with backgrounds in phonetics and speech processing, and ultimately received 81 responses. Figure~\ref{fig:attribute_edit} presents the results as a heatmap, where each column corresponds to one editing direction and each row corresponds to a voice attribute. The strong diagonal highlights confirm that each direction reliably changes the intended attribute, although some confusion arose between the last two dimensions.
\begin{figure}[!t]
\includegraphics[width=0.87\linewidth]{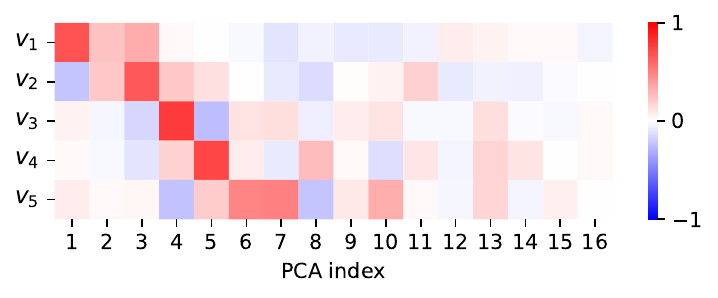}
\centering
\caption{Cosine similarity between the PCA directions used in the search algorithm and the five identified editing directions.}

\label{fig:cosinesim}
\vskip -0.2in
\end{figure}

Figure~\ref{fig:cosinesim} shows the cosine similarity between the PCA directions used in our search algorithm and the discovered voice editing directions. Despite originating from different methods, the PCA directions align closely with the editing directions, with each PCA direction primarily corresponds to one or two of the identified editing directions. This corroborates user feedback that many PCA directions are indeed interpretable, making the search process intuitive. We encourage readers to explore these patterns by visiting our demo page\footnote{https://myspeechprojects.github.io/voice-design-demo/}.

\section{Conclusions}
We presented a human-in-the-loop approach for creating personalized synthetic voices based on human feedback, aiming to help speech-impaired individuals regain their lost voices even if they cannot provide reference recordings. The approach leverages users' familiarity with their target voices, guiding them through an intuitive search process using listening and comparison tasks. Its effectiveness has been validated on various cases by computer simulations and user experiments. 

The quality of the search result, however, depends on the chosen starting point, particularly for out-of-domain voices. As a next step, we could investigate two potential methods to improve performance in these cases: (1) leveraging text-based voice generation models for initialization, and (2) developing human-in-the-loop algorithms for users to efficiently identify a close matching voice within a database for initialization.

This work focuses on English, leveraging the availability of a large, multi-speaker speech dataset to construct a latent speaker embedding space that represents diverse voices. However such extensive datasets are often lacking for lower-resource languages. We could extend this approach to multi-lingual settings, enabling similar voice design capabilities for languages with limited data resources.
\bibliographystyle{IEEEtran}
\bibliography{mybib}

\begin{thebibliography}{10}
\providecommand{\url}[1]{#1}
\csname url@samestyle\endcsname
\providecommand{\newblock}{\relax}
\providecommand{\bibinfo}[2]{#2}
\providecommand{\BIBentrySTDinterwordspacing}{\spaceskip=0pt\relax}
\providecommand{\BIBentryALTinterwordstretchfactor}{4}
\providecommand{\BIBentryALTinterwordspacing}{\spaceskip=\fontdimen2\font plus
\BIBentryALTinterwordstretchfactor\fontdimen3\font minus \fontdimen4\font\relax}
\providecommand{\BIBforeignlanguage}[2]{{%
\expandafter\ifx\csname l@#1\endcsname\relax
\typeout{** WARNING: IEEEtran.bst: No hyphenation pattern has been}%
\typeout{** loaded for the language `#1'. Using the pattern for}%
\typeout{** the default language instead.}%
\else
\language=\csname l@#1\endcsname
\fi
#2}}
\providecommand{\BIBdecl}{\relax}
\BIBdecl

\bibitem{veaux2012using}
C.~Veaux, J.~Yamagishi, and S.~King, ``Using hmm-based speech synthesis to reconstruct the voice of individuals with degenerative speech disorders,'' in \emph{Proc. Interspeech}, 2012, pp. 967--970.

\bibitem{erro2015personalized}
D.~Erro, I.~Hernaez, A.~Alonso, D.~Garc{\'\i}a-Lorenzo, E.~Navas, J.~Ye, H.~Arzelus, I.~Jauk, N.~Q. Hy, C.~Magarinos \emph{et~al.}, ``Personalized synthetic voices for speaking impaired: website and app.'' in \emph{Proc. Interspeech}, vol. 2015, 2015, pp. 1251--1254.

\bibitem{cave2021voicebank}
R.~Cave and S.~Bloch, ``Voice banking for people living with motor neurone disease: Views and expectations,'' \emph{Int. J. Lang. Commun. Disord.}, vol.~56, no.~1, pp. 116--129, 2021.

\bibitem{voiceclone}
S.~{\"{O}}. Arik, J.~Chen, K.~Peng, W.~Ping, and Y.~Zhou, ``Neural voice cloning with a few samples,'' in \emph{Annu. Conf. Neural Inf. Process. Syst.}, 2018, pp. 10\,040--10\,050.

\bibitem{svtotts}
Y.~Jia, Y.~Zhang, R.~J. Weiss, Q.~Wang, J.~Shen, F.~Ren, Z.~Chen, P.~Nguyen, R.~Pang, I.~L{\'{o}}pez{-}Moreno, and Y.~Wu, ``Transfer learning from speaker verification to multispeaker text-to-speech synthesis,'' in \emph{Annu. Conf. Neural Inf. Process. Syst.}, 2018, pp. 4485--4495.

\bibitem{adaspeech4}
Y.~Wu, X.~Tan, B.~Li, L.~He, S.~Zhao, R.~Song, T.~Qin, and T.~Liu, ``Adaspeech 4: Adaptive text to speech in zero-shot scenarios,'' in \emph{Proc. Interspeech}, 2022, pp. 2568--2572.

\bibitem{naturalspeech3}
Z.~Ju, Y.~Wang, K.~Shen, X.~Tan, D.~Xin, D.~Yang, Y.~Liu, Y.~Leng, K.~Song, S.~Tang, Z.~Wu, T.~Qin, X.-Y. Li, W.~Ye, S.~Zhang, J.~Bian, L.~He, J.~Li, and S.~Zhao, ``Naturalspeech 3: zero-shot speech synthesis with factorized codec and diffusion models,'' in \emph{Proc. ICML}, 2024.

\bibitem{sounddesign}
E.~R. Miranda and E.~R. Miranda, \emph{\BIBforeignlanguage{eng}{Computer sound design : synthesis techniques and programming}}, 2nd~ed., ser. Music technology series.\hskip 1em plus 0.5em minus 0.4em\relax Oxford: Focal Press, 2002.

\bibitem{prompttts}
Z.~Guo, Y.~Leng, Y.~Wu, S.~Zhao, and X.~Tan, ``Prompttts: Controllable text-to-speech with text descriptions,'' in \emph{Proc. ICASSP}, 2023, pp. 1--5.

\bibitem{leng2024prompttts2}
Y.~Leng, Z.~Guo, K.~Shen, Z.~Ju, X.~Tan, E.~Liu, Y.~Liu, D.~Yang, leying zhang, K.~Song, L.~He, X.~Li, sheng zhao, T.~Qin, and J.~Bian, ``Prompt{TTS} 2: Describing and generating voices with text prompt,'' in \emph{Proc. ICLR}, 2024.

\bibitem{prompttts++}
R.~Shimizu, R.~Yamamoto, M.~Kawamura, Y.~Shirahata, H.~Doi, T.~Komatsu, and K.~Tachibana, ``Prompttts++: Controlling speaker identity in prompt-based text-to-speech using natural language descriptions,'' in \emph{Proc. ICASSP}, 2024, pp. 12\,672--12\,676.

\bibitem{promptspeaker}
Y.~Zhang, G.~Liu, Y.~Lei, Y.~Chen, H.~Yin, L.~Xie, and Z.~Li, ``Promptspeaker: Speaker generation based on text descriptions,'' in \emph{Proc. ASRU}, 2023, pp. 1--7.

\bibitem{vyas2023audiobox}
A.~Vyas, B.~Shi, M.~Le, A.~Tjandra, Y.-C. Wu, B.~Guo, J.~Zhang, X.~Zhang, R.~Adkins, W.~Ngan \emph{et~al.}, ``Audiobox: Unified audio generation with natural language prompts,'' \emph{arXiv preprint arXiv:2312.15821}, 2023.

\bibitem{hai2024dreamvoice}
J.~Hai, K.~Thakkar, H.~Wang, Z.~Qin, and M.~Elhilali, ``Dreamvoice: Text-guided voice conversion,'' in \emph{Proc. Interspeech}, 2024, pp. 4373--4377.

\bibitem{lyth2024natural}
D.~Lyth and S.~King, ``Natural language guidance of high-fidelity text-to-speech with synthetic annotations,'' \emph{arXiv preprint arXiv:2402.01912}, 2024.

\bibitem{voiceme}
P.~van Rijn, S.~Mertes, D.~Schiller, P.~Dura, H.~Siuzdak, P.~M.~C. Harrison, E.~Andr{\'{e}}, and N.~Jacoby, ``Voiceme: Personalized voice generation in {TTS},'' in \emph{Proc. Interspeech}, 2022, pp. 2588--2592.

\bibitem{robotvoice}
P.~van Rijn, S.~Mertes, K.~Janowski, K.~Weitz, N.~Jacoby, and E.~Andr{\'{e}}, ``Giving robots a voice: Human-in-the-loop voice creation and open-ended labeling,'' in \emph{Proc. CHI Conf. Human Factors Comput. Syst.}, 2024, pp. 584:1--584:34.

\bibitem{udagawa22_interspeech}
K.~Udagawa, Y.~Saito, and H.~Saruwatari, ``Human-in-the-loop speaker adaptation for dnn-based multi-speaker tts,'' in \emph{Proc. Interspeech}, 2022, pp. 2968--2972.

\bibitem{hilvoice}
X.~Chen, Q.~Huang, X.~Wu, Z.~Wu, and H.~Meng, ``Hilvoice:human-in-the-loop style selection for elder-facing speech synthesis,'' in \emph{Proc. ISCSLP}, 2022, pp. 86--90.

\bibitem{nansy}
H.~Choi, J.~Lee, W.~Kim, J.~Lee, H.~Heo, and K.~Lee, ``Neural analysis and synthesis: Reconstructing speech from self-supervised representations,'' in \emph{Annu. Conf. Neural Inf. Process. Syst.}, 2021, pp. 16\,251--16\,265.

\bibitem{wright2015coordinate}
S.~J. Wright, ``Coordinate descent algorithms,'' \emph{Math. Program.}, vol. 151, no.~1, pp. 3--34, 2015.

\bibitem{dio}
M.~Morise, H.~Kawahara, and H.~Katayose, ``Fast and reliable f0 estimation method based on the period extraction of vocal fold vibration of singing voice and speech,'' in \emph{Int. Conf. Audio Eng. Soc.}\hskip 1em plus 0.5em minus 0.4em\relax Audio Engineering Society, 2009.

\bibitem{ecapa-tdnn}
B.~Desplanques, J.~Thienpondt, and K.~Demuynck, ``{ECAPA-TDNN:} emphasized channel attention, propagation and aggregation in {TDNN} based speaker verification,'' in \emph{Proc. Interspeech}, 2020, pp. 3830--3834.

\bibitem{contentvec}
K.~Qian, Y.~Zhang, H.~Gao, J.~Ni, C.~Lai, D.~D. Cox, M.~Hasegawa{-}Johnson, and S.~Chang, ``Contentvec: An improved self-supervised speech representation by disentangling speakers,'' in \emph{Proc. ICML}, vol. 162, 2022, pp. 18\,003--18\,017.

\bibitem{libritts-r}
Y.~Koizumi, H.~Zen, S.~Karita, Y.~Ding, K.~Yatabe, N.~Morioka, M.~Bacchiani, Y.~Zhang, W.~Han, and A.~Bapna, ``Libritts-r: {A} restored multi-speaker text-to-speech corpus,'' in \emph{Proc. Interspeech}, 2023, pp. 5496--5500.

\bibitem{vits}
J.~Kim, J.~Kong, and J.~Son, ``Conditional variational autoencoder with adversarial learning for end-to-end text-to-speech,'' in \emph{Proc. ICML}, vol. 139, 2021, pp. 5530--5540.

\bibitem{vctk}
J.~Yamagishi, C.~Veaux, S.~King, and S.~Renals, ``Speech synthesis technologies for individuals with vocal disabilities: Voice banking and reconstruction,'' \emph{Acoust. Sci. Technol.}, vol.~33, no.~1, pp. 1--5, 2012.

\bibitem{spectralregularizer}
A.~Ramesh, Y.~Choi, and Y.~LeCun, ``A spectral regularizer for unsupervised disentanglement,'' \emph{CoRR}, vol. abs/1812.01161, 2018.

\bibitem{lowranksubspace}
J.~Zhu, R.~Feng, Y.~Shen, D.~Zhao, Z.~Zha, J.~Zhou, and Q.~Chen, ``Low-rank subspaces in gans,'' in \emph{Annu. Conf. Neural Inf. Process. Syst.}, 2021, pp. 16\,648--16\,658.

\bibitem{closedform}
Y.~Shen and B.~Zhou, ``Closed-form factorization of latent semantics in gans,'' in \emph{Proc. CVPR}, 2021, pp. 1532--1540.

\bibitem{jacobianregularizer}
Y.~Wei, Y.~Shi, X.~Liu, Z.~Ji, Y.~Gao, Z.~Wu, and W.~Zuo, ``Orthogonal jacobian regularization for unsupervised disentanglement in image generation,'' in \emph{Proc. ICCV}, 2021, pp. 6701--6710.

\bibitem{dbscan}
M.~Ester, H.~Kriegel, J.~Sander, and X.~Xu, ``A density-based algorithm for discovering clusters in large spatial databases with noise,'' in \emph{Proc. Int. Conf. Knowl. Discovery Data Mining}, 1996, pp. 226--231.

\end{thebibliography}

\end{document}